# A review of information security aspects of the emerging COVID-19 contact tracing mobile phone applications

Georgios Magklaras[1,2] https://orcid.org/0000-0001-5836-8252 and Lucia N. López-Bojórquez[2]

[1] Norwegian Center for Molecular Medicine, University of Oslo, Norway and Steelcyber Scientific, Oslo, Norway

[2] Steelcyber Scientific, Oslo, Norway
georgios@steelcyber.com

**Abstract.** This paper discusses the aspects of data reliability and user privacy for the emerging practice of mobile phone based contact tracing for the COVID-19 pandemic. Various countries and large technology companies have already used or plan to design and use mobile phone based solutions, in an effort to urgently expedite the process of identifying people who may have been exposed to the disease and limit its spread to the general population. However, serious concerns have been raised both in terms of the validity of the collected data as well as the extent to which implemented approaches can breach the privacy of the mobile phone users. This review examines the weaknesses of existing implementations and concludes with specific recommendations that can contribute towards increasing the safety of infrastructures that collect and process this kind of information, as well as the adoption and acceptance of these solutions from the public.

## 1      Introduction

On March 11[th] 2020, the Director General of the World Health Organization (WHO) declared the outbreak of COVID-19 a global pandemic [1]. Emergency measures have fundamentally altered the economy and society on a global scale, as health systems around the world struggled to keep up with the demand for emergency health care [2]. As part of these measures and in an attempt to quickly identify people who may have been exposed to the disease and thus limit its spread to the general population, many governments around the world have deployed mobile phone applications to make the public health process of contact tracing more efficient in a massive scale. A non exhaustive list of countries that were among the first to deploy mobile phone based contract tracing applications include Australia [3], China [4], Israel [5], Norway [6], Singapore [7] and South Korea [8]. In addition, large technology companies such as Google and Apple are preparing their own infrastructure for COVID-19 contact tracing [9].

Many of the previously mentioned governments that were early adopters of the technology and make the participation of its citizens in electronic contract tracing voluntary have claimed that their applications are safe to use and prompted their citizens to download and use them. However, many technology experts have criticized the technology [10] or expressed concern about its efficacy versus its privacy implications [11]. Moreover, in certain countries, public response to the technology was lukewarm. For instance, India, Singapore and Norway have seen limited user acceptance of these solutions if one examines recently estimated application download numbers [12]. All these facts give merit to a closer examination of the problems of COVID-19 contact tracing solutions.

Before taking a closer look into the problems of contact tracing solutions, it is necessary to provide essential definitions about the concept and the technologies involved in making the transition from manual to electronic procedures.

In public health epidemiological context, contact tracing is the process of identifying persons who may have come into contact with a person whose infection has been confirmed [13]. The infected person is often referred to as the "index case" and all the people that have come into contact that meets certain criteria (proximity, type of transmission, duration) with the index case are referred to as the "contacts". The systemic collection of further information about these contacts aims to isolate them, test them for the infection and treat them where applicable. Depending on the type and expected spread of an outbreak, the process can be recursively repeated for contacts of contacts. The overall aim is to limit the spread of the infection in the general population.

Health authorities follow specific protocols that require manual contact tracing. This means that health workers evaluate the provided information, search for locating the contacts, notify the contacts (phone call) and all this depends on the accuracy of the information that the index case and his/her subsequent contacts can provide. It is thus reasonable to assume that as health infrastructures are strained for resources in a fast spreading infection, the quality as well as the accuracy of manual contact tracing procedure will suffer. This has been confirmed well before the COVID-19 outbreak. In fact, electronic contact tracing has been tested in the pre-COVID-19 world in many epidemiological emergencies, among them the Ebola virus outbreak [14]. Although this study is far from the technology implementation aspects we see in the COVID-19 mobile phone contact tracing solutions, it highlighted the power of the ubiquity of the mobile phone as a tool to aid the monitoring and spread of infectious diseases.

In a post-COVID-19 world, governments and technology companies turn to various aspects of mobile and general computing infrastructures to implement contact tracing solutions. In particular, most COVID-19 contact tracing solutions make use of the following mobile phone technologies:

A) The use of Global Positioning System and Assisted GPS (A-GPS) [15] technology: Every mobile phone has an embedded GPS receiver and through to complimentary components of a 3GPP compliant [16,17] telecommunications infrastructure, a time series of GPS coordinates of the mobile device can be recorded. Features like the Google Account Location history [18], as well as the Chinese [4], Israeli [5] and Norwegian [6] contact tracing applications make use of the position/location data. Google has also used location data during the COVID-19 pandemic to estimate the extent of the imposed quarantine measures in various countries with the so called 'mobility reports' [19].

B) The use of the Bluetooth protocol [20]: The Bluetooth protocol is a complex wireless technology standard that encompasses different modes of transmission and functionality. The relevant bits to contact tracing concern its low energy variant called Bluetooth LE [21]. This variant is used to perform proximity sensing calculations. The calculations are used to estimate the distance between the index case and the contacts and thus play a crucial role in most COVID-19 contact tracing application implementations. Another crucial aspect that concerns the Bluetooth operation is that the technology is used to exchange data between devices. Latter paragraphs will describe that process in more detail.

C) The increase in power and data storage in mobile phones, as well as the ubiquity of reliable 3G/4G (and in the near future 5G) connections create powerful ways of constructing big data sets with different levels of anonymity and susceptibility to linkage attacks [22]. Most of the solutions claim that they take precautions to anonymize the data they exchange. Data exchange and collection can also occur in de-centralized or centralized ways. This has different implications for the privacy of the users that contribute the data in question.

Leaving the substantial variations among existing different COVID-19 contact tracing implementations to the side, in simple terms, when a user downloads a contact tracing application to a smartphone, the device will in principle perform the following actions:

A)Activate the Bluetooth LE interface and will broadcast its presence by means of transmitting an anonymous identifier. The transmission of the identifier is performed repeatedly in the form of a beacon.

B)Use the same Bluetooth LE interface to record received anonymous identifiers of other mobile phones within range.

C)For every received/intercepted anonymous Bluetooth LE identifier, the phone will attempt to estimate its proximity. This proximity sensing step is crucial to the validity of the sampled data.

D)The collected data are stored in the smartphone but are handled in different ways. An abstraction of such a record could look like the ones below:

$$r:time,date,BLE\_id,proximity\_estimation,covid19\_flag$$
$$\text{OR}$$
$$r:time,date,BLE\_id,proximity\_estimation,A\_GPS\_data,covid19\_flag$$

where BLE_ID is the anonymous identifier, proximity_estimation represents a distance (meters) , A_GPS_data represent location data of the smartphone according to the data collected by its A-GPS receiver and finally covid19_flag represents whether the user of the smartphone has disclosed (voluntarily) whether he is infected with COVID-19. Different contact tracing implementations upload these records (with the user's consent) to different types of central database infrastructures for processing.

For the purposes of clarity, we need to emphasize not all mobile application implementations collect GPS data (A_GPS_data field). The collection of location data creates privacy concerns that are discussed in Section 3 of this paper. The A_GPS_data field can collect other forms of location data (Cell tower ID) to aid the accuracy of the proximity sensing process in various ways.

A central database will process the collected records with particular emphasis on the records that have the covid19_flag set and the proximity_estimation within a certain range (say for instance less than 2 meters or less). Consequently, it is possible to message alert all users that have been within a pre-defined proximity and time exposure of a specific BLE_id whose smartphone user has declared his/her infection.

It is therefore evident that smartphones can provide time, location and proximity data that public health authorities consider valuable, in order to alert the general population [23]. This process forms the very basis of smartphone based COVID-19 contact tracing and will be used as a reference mechanism for analysis for the rest of this paper.

The following sections will focus on various implementation details of the reference mechanism. Section 2 will discuss information security aspects that concern the use of the Bluetooth LE protocol, its data accuracy, as well as its various information security weaknesses. Section 3 elaborates on the privacy aspects of storing anonymous data in central infrastructures. The fourth and final section of the paper concludes with concrete recommendations that aim to improve the security of electronic contact tracing solutions.

## 2      Bluetooth LE issues and contact tracing

The Bluetooth protocol is a vast and complex specification [20]. Different versions and smartphone chipset implementations can result in different operational and information security aspects of its use for the purposes of contact tracing. However, in broad terms, these aspects touch on three different areas. The first is the area of user privacy. One needs to question what is the likelihood that a user can be identified as a result of the Bluetooth data exchange necessary to facilitate contact tracing. A second question relates to how accurate are the data collected by Bluetooth LE for the purposes of contact tracing. Finally, a third question to raise is what are the security implications of using it to broadcast your (in theory anonymous) presence and exchange data with devices you do not know.

Bluetooth LE allows device manufacturers to use temporary random addresses in over-the-air communication instead of their permanent address to prevent tracking, as part of the Bluetooth Core Specification version 4 [24, 25]. Earlier versions of the Bluetooth Core Specification were broadcasting the interface MAC address, a permanent identifier that is unique for every smartphone [26] and could thus be used to track an individual. While Bluetooth Core Specification version 4 addresses this issue, it also leaves gaps that could be exploited and lead, under specific circumstances, to identification of individuals.

Jameel and Dungen [27] examined Bluetooth LE beacon protocols and an array of mechanisms that facilitate localized interactions with smartphones and other Bluetooth devices via the beacon mechanisms. The advlib library [28] is a product of their work which allows software developers to easily integrate Bluetooth LE beacon advertising-based functionality into their applications, without having to embed them into the low-level protocol mechanisms. However, the practical application of this work for an adversary is that the library could be used to identify Bluetooth powered devices. While it is not possible to track a specific individual by making use of this mechanism, identifying that someone has a specific phone and a specific accessory in an area with a limited number of people could aid the process of adversarial reconnaissance aiming towards personal identification.

Becker, *et al* [29] proceed further and demonstrate that even current Bluetooth LE anonymization measures are vulnerable to passive tracking. Their work proposes an address-carryover algorithm that exploits the asynchronous nature of the Bluetooth LE payload and achieves tracking that bypasses the attempted address randomization of a device. The worrying aspect of their study and experimental setup is that it does not use differential cryptanalysis to decrypt the content of Bluetooth LE communication. Their method works entirely by intercepting public, unencrypted Bluetooth LE advertising traffic which is necessary for steps A and B of the abstracted COVID-19 contact tracing procedure outlined in Section 1 of this paper. It is broad, in the sense that it is effective against all iOS, macOS and Windows 10 devices.

Another worrying aspect of the work outlined in [29] and also supported by other theoretical and experimental work [30,31] is that despite the existence of Bluetooth MAC address randomization mechanisms to achieve anonymity, not all device manufacturers and operating system/application authors choose to employ them in the same way. There is a certain amount of flexibility in how to implement and transmit these randomized identifiers. These might include standard ways but different operating systems and applications might embed additional information as part of the Bluetooth LE public beacon payloads for the purposes of incorporating customized functionality. This additional information often leaks vital identity aspects and is dictated by software, from the operating system all the way to the application layer. Consequently, different COVID-19 contact tracing applications diverge substantially from whatever the relevant Bluetooth standards dictate and offer different levels of user privacy.

As far as the data accuracy of Bluetooth LE collected data is concerned, there are also serious doubts expressed by experts. Step C of the abstracted COVID-19 contact tracing procedure (Section 1 of this paper) attempts to estimate the distance of an intercepted Bluetooth LE beacon. The question here is with what accuracy can Bluetooth LE determine whether the user of another smartphone is closer than a predetermined distance (say 2 meters). The best way to answer that question is to understand the mechanism employed to measure that distance.

The Bluetooth protocol uses the Received Signal Strength Indicator (RSSI) to measure distance between devices [32]. The principle is that the stronger the signal, the closer the devices are to each other, so a correlation between sensed signal strength and distance can be approximated. However, different bluetooth chipset implementations utilize the RSSI in slightly different ways. While appropriate calibration can reduce these inaccuracies, the problems do not stop there. The Bluetooth LE transmission frequency often interferes with other devices in the 2.4 GHz range, such as older WiFi routers, unshielded USB cables and microwave ovens. A Bluetooth LE device would do its best to extend the 'beacons' (advertisement of presence and availability) by keeping constant time and regulating the transmission power to overcome other sources of interference. In such a frequency congested environment, a real distance of 1.5 meters could really be estimated as 2.5 meters (false negative), or a real distance of 2.5 meters could be  estimated to under 1.5 meters (false positive). Many experts, amongst them the Bluetooth inventors Jaap Haartsen and Sven Mattisson, agree that these proximity sensing inaccuracies were and remain a limiting factor [33]. As a result, the accuracy of the collected proximity data will be reduced and further post processing steps are needed, in order to allow someone to derive safe conclusions about who is in real danger to get infected due to proximity.

Finally, an often overlooked aspect of Bluetooth LE is its transmission range. While Bluetooth LE version 4 has a Line Of Sight (LOS) beacon range of 430 meters, the next major version of the protocol specification (v5) extends that LOS range to 780 meters [34]. At the time of writing, most mobile phones will be supporting Bluetooth LE version 5 within the next 12 months. If every smartphone used to perform many personal and business critical things (e-banking, remote control of systems at work, email) has yet another interface that advertises the presence of an individual (apart from the 4G/5G and WiFi interfaces), this provides an advantage for an adversary and can act as a catalyst for cyberattack vectors. The fact is that bluesnarfing attacks against mobile phones have been identified from the early adoption days of the bluetooth protocol [35]. Moreover, there is good evidence that these attacks have persisted over a number of years [36] and will continue to persist with many recent notable examples that target bluetooth device firmware features [37, 38]. The conclusion derived by this body of work is that the COVID-19 contact tracing applications increase the exploitable attack footprint of the average smartphone.

## 3    Privacy and security aspects of storing and processing contact tracing data

The COVID-19 contact tracing data collected by smartphones always require some data entry processing backend (central server or servers that operate independently). However, there are different degrees of data centralization among the various solutions. For instance, the Norwegian [6] and Singaporean [7] contact tracing implementation are some of the paradigms that require all collected data to be centralized for further processing. In direct contrast, the Temporary Contact Numbers (TCN) protocol [39] as well as the Decentralized Privacy-Preserving Proximity Tracing (DP-3T) protocol [40] constitute examples of protocols that are designed to minimize both the amount of info as well as the necessary processing in a centralized infrastructure. Google and Apple seem to follow the decentralized approach [9].

Prior discussing the relative merits of centralized versus decentralized COVID-19 contact tracing approaches and beyond the Bluetooth LE related privacy threats discussed in Section 2, it is useful to examine the context of what user privacy means when combined with a justified need to enhance the tools that health authorities can utilize to effectively contain the spread of a pandemic.

The European Union is among the major global players that have officially recognized the potential of smartphone and associated technological solutions to fight the COVID-19 pandemic [41]. Part of this recognition is made amidst the presence of comprehensive regulations such as GDPR [42] that set very strict requirements for the storage and processing of personal information. Many countries have modified their national data protection laws to make urgent allowances for the data collection and processing of personal data related to the COVID-19 pandemic [43]. As an example, the Norwegian National Data Protection Authority (Datatilsynet) has explicitly permitted non anonymous location data processing for the purposes of COVID-19 smartphone contact tracing, only if it is not possible to derive safe conclusions from anonymous proximity based data [44]. These steps indicate that there is a need for balance between personal privacy and public health [45].

It is outside the scope of this paper to pass a judgement on whether amendments to national legislations should favor privacy over public health or vice versa. The goal of this review is to highlight what is in favor of the privacy of the smartphone user and thus help specialists and policy makers to implement electronic contact tracing in the least privacy intrusive manner. Achieving such a goal is not always trivial and it will require adherence to international standards. Validated international standards for smartphone based contact tracing do not exist at the time of writing. What does exist is a set of EU recommendations [46] that dictate a set of principles relevant to user privacy in the context of electronic contact tracing. In particular, the EU recommendations dictate that all smartphone based contact tracing solutions should:

A)Operate on anonymized data with the goal of alerting users that have been in close proximity to confirmed cases without revealing the identity of the index case or the contacts. Breach of anonymity and hence disclosure of the identity details of an individual

B) Not track the location of the users.

C) Be based on voluntary user participation. Any unauthorized usage of data without the knowledge or the approval of the user is strictly prohibited.

D)The entire infrastructure should be secure and effective end to end. This includes any centralized components where data are deposited for processing.

E) There need to be interoperable and scalable across a number of countries, as people travel from country to country.

Having these requirements as a guide, one of the first conclusions we can derive is that any solution that stores, sends and processes GPS and A-GPS data is not acceptable from a privacy perspective. A time series of GPS coordinates or other network assisted location data (cell tower ID) is personal information and whether deposited partly or completely in a central database server reveals too much information for a user. Research efforts that propose privacy preserving location based contact tracing exist. MIT researchers have proposed a contact tracing system based on a method that redacts, transforms and encrypts GPS coordinates to address the privacy preservation problem [47]. The contact tracing is then computed by a process known as Private Set Intersection (PSI), a technique commonly employed as part of secure multiparty computing [48], aiming to reveal only the common data values that are necessary for the computation.

However, privacy preserving contact tracing techniques that use GPS coordinates constitute best effort experimental approaches that need a reference implementation to be tested and proven. An additional practical matter is that of accuracy. GPS and A-GPS coordinates cannot at the moment provide a level of accuracy in terms of contact proximity and this is why most solutions today resort to the use of Bluetooth LE, even with the problems discussed in Section 2 of this paper. A last practical aspect concerns compliance to existing legislation. If the law does not provide a clear framework for the sampling of location data for health related purposes, then it is not possible to employ these techniques and thus approaches that rely on geolocating the users will be impractical and impossible to implement.

Researchers that are proponents of techniques that employ GPS coordinates [47] point out that large companies already collect user location data for operational and advertisement purposes. While this is true [18,19], there is a distinct difference between geolocating individuals for commercial purposes and doing the same in a health context. Apart from the location info this kind of contact tracing solutions contain references to health status (infected or not infected status of an individual). Combining personal location info to health status raises the legal context and regulatory handling requirements of the collected/processed information. For instance, the European Data Protection Supervisor considers all data concerning health as a special category [49,50] for which strict privacy preserving requirements apply when it comes to the handling and processing of the collected information.

The concentration of large amounts of (theoretically) anonymous health related information in central repositories for the purposes of centralized contact tracing solutions [6,7] creates certain risks and operational requirements for the storage and processing of the data. Weaknesses in the anonymity protocols (such as the ones described in Section 2 of this paper in connection to the Bluetooth LE protocol) or in the implementation of infrastructures could place a malicious adversary in a situation to collect information that could compromise the privacy of millions of individuals. The handling of large amounts of anonymous (or desensitized) health data predates the electronic contact tracing era and can be observed in other fields of health informatics. A good example is that of genomic medicine where certain types of genomic data, even if they have been anonymised in principle, they do provide distinct probable ways to re-identify the subjects of a study [51]. For these reasons, access to these types of data requires data consumers to follow certain ethical guidelines that bind them not to use them in ways that could re-identify the anonymised study subjects and conform to strict storage and data processing requirements [52].

On the other hand, centralized processing requirements are simpler to implement in principle when compared to decentralized contact tracing solutions such as those proposed in [39] and [40]. In general terms, the aim of decentralized contract tracing solutions is to reduce the privacy and security impact of having all the necessary data in one place. They still require a minimal centralized component, especially for steps that incorporate the health status (infected or not infected contact), however the disclosure of information to central entities is minimal by design. This reduces the possibility for abuse of central data repositories. On the other hand, decentralized solutions delegate the processing of info to non trusted devices (the smartphones of the users). This increases implementation complexity. The entire concept has not been yet proven at scale, both in theory and practice. Most existing contact tracing solutions follow the centralized storage and information processing model at the time of writing.

A final consideration has to do with how the central IT infrastructure for contract tracing solutions are implemented. There seems to be certain lack of transparency on how this central part has been implemented. Taking Norway as an example, a country with good tradition on respecting the privacy of its citizens and among the first to launch a COVID-19 contact tracing application, it is evident that no tender processes have been disclosed for awarding public funds to construct the application [53], calls to open source the application in order to aid the review by security experts were denied [54] and that data that contain GPS, Bluetooth LE smartphone identifiers and health status were stored in private cloud vendors [55] with unclear status on whether the data can leave the Norwegian geographic border. As a result, the Norwegian implementation drew a lot of criticism by many IT experts around the world [56]. This is by no means unique to Norway. Other countries have faced similar criticism.

Transparency of data processing, as well as export control of health data are issues that should be taken seriously as dictated by pan European (GDPR) and other international legislation [57]. Besides compliance, choices that limit transparency make public acceptance of a technology difficult. Thus, it is evident that implementing contact tracing technology should be a process with structure and best practices that are missing at the moment. This structure and recommended practices forms the subject of the next section of this paper.

# 4 Conclusions and recommendations for implementing electronic contact tracing solutions

The previous sections of this paper have highlighted that the existing COVID-19 contact tracing applications have serious problems, both in terms of the reliability of the collected data sets, as well as in terms of preserving the end user privacy and security. Addressing these problems is not a trivial process and will require substantial efforts towards the creation of standards that oversee the development of contact tracing platforms. The existing EU recommendations [46] that were discussed in Section 3 of this paper can serve as a good start on a road map that will make electronic contact tracing both usable and acceptable by societies around the world.

On the issues of Bluetooth LE accuracy discussed in Section 2 [33], there are research and development approaches aiming to increase the proximity sensing accuracy of the protocol. Examples of such work can be found in [58-59]. It is also possible that smartphone chipset manufacturers together with future versions of the Bluetooth LE protocol will add features that will increase the proximity sensing accuracy. However, no matter what technological measures are employed to achieve additional proximity sensing precision, the important thing is to put them to the test in a standard manner. The only reliable way to do this is to set control experiments where a group of individuals using smartphones can create verified/predetermined contacts under a variety of conditions (inside buildings with different level of RF noise environments different contact times and different number of individuals). If the subsequent analysis of the recorded data accurately represents the verified/predetermined conditions within a predetermined statistical accuracy (say less than 1% for both false positives or negatives) then this means that the data collected by a contact tracing implementation is good enough to be used for the public. Launching an application on a national scale without proving the accuracy of the sampled data and verifying it by statisticians and experts can lead to misleading results and should be avoided.

When it comes to the rest of the vulnerabilities of the Bluetooth LE protocol (range on LOS and software vulnerabilities discussed in Section 2 and referenced in [34-38]), there are various measures to be taken. It is prudent that the Bluetooth LE power is regulated in a standardized manner when operating a contact tracing application, so that the effective range of the protocol is reduced. Setting devices to the lowest power level to perform reliably proximity sensing will reduce the effective adversarial surveillance range [60]. In addition, smartphone manufacturers need to do a better job in addressing the firmware and mobile operating system vulnerabilities, especially for the older smartphone devices. As an example, in the Android mobile operating system, critical Bluetooth vulnerabilities such as the 'BlueFrag' CVE-2020-0022 [61] affected mainly older versions of the Android system for several months. While the vulnerability in question has been patched at the time of writing, not all Android device manufacturers have included this patch in their Android OEM versions. The result is that a substantial number of smartphone users that still operate Android version 8 are vulnerable if they use contact tracing and other Bluetooth based data exchange applications. Thus, it is our view that world wide or regional regulations should make mandatory that all smartphone vendors issue critical system updates throughout the expected life cycle of a smartphone (3-5 years).

Drawing upon the EU contact tracing implementation requirements [46], we advise against the usage of any location data (GPS, A-GPS, cell tower ID or other) in electronic contact tracing solutions. Apart from conflicts with data protection legislation discussed in Section 3 [49, 50], we do not see how location data can enhance the contact discovery. For the purposes of contact tracing, the Bluetooth LE proximity collected data are more relevant and accurate than any other form of satellite or network assisted location system. Incorporating location data, even when anonymised/desensitized increases the susceptibility of the collected data to differential privacy attacks [62], especially in implementations where the data is centralized and should be avoided.

We do not have enough data on existing implementations to recommend whether existing decentralized approaches should be favored over centralized approaches. As discussed in Section 3 of this paper there are certain advantages and disadvantages for each of these approaches. Decentralized approaches follow the principle of minimizing the amount of information necessary to perform the contact tracing, however they add implementation complexity and require information to be distributed to untrusted entities. While decentralized approaches look promising, they require further theoretical and practical implementation validation by experts, before definite conclusions are drawn. However, as both approaches require some main IT infrastructure component beyond the information gathered by smartphones, the following paragraphs discuss some concrete recommendations that can aid the security of electronic contact tracing solutions.

Section 3 discussed the paradigm of genomic medicine data [51] and its analogy to that of electronic contact tracing solutions. The common denominator is the presence of a large amount of anonymized health data. Whatever cryptographic precautions can be taken to protect the identity of the contact tracing users, this does not change the fact that a large amount of information about public health is stored in one form or another (centralized versus decentralized, different encryption standards). In our view, this should be good enough to treat this kind of anonymous data in the same way as eponymous medical data. This view is supported by existing data classification policies that form part of Information Security Management practices [63]. As an example, the University of Oslo, the largest and oldest academic institution in Norway, manages large amounts of electronic information, including sensitive eponymous data from the Oslo University Hospital. For that reason, its information security management system [64] classifies large amounts of anonymous health data at the highest level of data sensitivity [65]. This has several implications about how anonymised contract tracing information should be stored and processed.

Infrastructures that hold eponymous sensitive medical data and have approval by relevant national data protection authorities implement a lot of technical requirements to ensure that the confidentiality, integrity and availability of the sensitive data is safeguarded. Drawing from the University of Oslo's paradigm, its 'Services for Sensitive Data (TSD)' platform [66] is a practical implementation that provides these safeguards. Elements such as multi-factor authentication [67], compartmentalization of computation activities on security hardened virtual machines and storage/backup encryption are some of the techniques employed by TSD. In our view, these should be mandatory technical elements that should form a standard for every core IT infrastructure platform that handles electronic contact tracing data at national/international level. In addition, core IT infrastructures should comply to GDPR [42] and possibly the HIPAA standard [68]. Compliance to these standards can also aid the interoperability among different national contact tracing solutions across a number of countries and continents. EU requirements dictate that contact tracing solutions should be interoperable [46].

Finally, as the use of cloud computing is increasing and the pressure for healthcare systems to be more cost effective is growing [69], there are certain risks associated to placing public health data in the cloud. A principal risk is that many large private cloud providers offer a utility service without safeguarding (or even wanting to know) the criticality and importance of the data and the tasks performed in their infrastructure [70]. When private cloud providers are used for core IT contact tracing infrastructure, we recommend three concrete rules. The first is that private cloud providers should comply to the same technical requirements and regulations set of the previous paragraph. In addition and as a consequence of regulatory compliance, private cloud providers should provide IT infrastructures within the geographical territory of the country/region if laws dictate the data should be localized. A third recommendation is that an independent cost-risk analysis should be commissioned prior reaching decisions to store and process contact tracing data exclusively in private cloud providers. A better approach is to adopt hybrid cloud technologies, where a public authority can have the option of easily turning the data and compute activities back to their own infrastructure, in case they face legislation or data availability problems.